\documentclass[onecolumn]{autart} 

\usepackage{amsmath,amssymb,mathtools}
\usepackage{enumitem}
\usepackage{cancel}
\usepackage{subcaption}
\usepackage{xcolor}
\usepackage{graphicx} 

\begin{document}

\begin{frontmatter}

\title{A Nonlinear Perron-Frobenius Approach \\ for Stability and Consensus \\ of Discrete-Time Multi-Agent Systems\thanksref{footnoteinfo}} 

\thanks[footnoteinfo]{This work was supported in part by the Italian Ministry of Research and Education (MIUR) with the grant ``CoNetDomeSys", code RBSI14OF6H, under call SIR 2014 and by Region Sardinia (RAS) with project MOSIMA, RASSR05871, FSC 2014-2020, Annualita' 2017, Area Tematica 3, Linea d'Azione 3.1.}

\author[Cagliari]{Diego Deplano}\ead{diego.deplano@diee.unica.it}, 
\author[Cagliari]{Mauro Franceschelli}\ead{mauro.franceschelli@diee.unica.it}, 
\author[Cagliari]{Alessandro Giua}\ead{giua@unica.it} 

\address[Cagliari]{Department of Electrical and Electronic Engineering, University of Cagliari, Italy} 


\begin{abstract} 
In this paper we propose a novel method to establish stability and, in addition, convergence to a consensus state for a class of discrete-time Multi-Agent System (MAS) evolving according to nonlinear heterogeneous local interaction rules which is not based on Lyapunov function arguments. In particular, we focus on a class of discrete-time MASs whose global dynamics can be represented by sub-homogeneous and order-preserving nonlinear maps. This paper directly generalizes results for sub-homogeneous and order-preserving linear maps which are shown to be the counterpart to stochastic matrices thanks to nonlinear Perron-Frobenius theory. We provide sufficient conditions on the structure of local interaction rules among agents to establish convergence to a fixed point and study the consensus problem in this generalized framework as a particular case. Examples to show the effectiveness of the method are provided to corroborate the theoretical analysis.
\end{abstract}

\end{frontmatter}

\pagebreak

\section{Introduction}

	The study of complex systems where local interactions between individuals give rise to a global collective behavior has aroused much interest in the control community. Such complex systems are often called Multi-Agent Systems (MAS), consisting of multiple interacting agents with mutual interactions among them.
	A topic that captured the attention of many researchers is the consensus problem \cite{Olfati2007}, where the objective is to design local interaction rules among agents such that their state variables converge to the same value, the so called agreement or consensus state.
	
	A MAS can be modeled as a dynamical system. In the discrete time linear case, classical Perron-Frobenius Theory is crucial in the convergence analysis. Indeed, in one of the most popular works in this topic (\cite{jadbabaie2003coordination}), the authors established criteria for convergence to a consensus state for MAS whose global dynamics can be represented by linear time-varying systems with non-negative row-stochastic state transition matrices, which are object of study of the classical Perron-Frobenius Theory. 	
	The notable aspect of this work was to exploit such theory and graph theory instead of Lyapunov theory, allowing to study systems for which finding a common Lyapunov function to establish convergence is difficult or even impossible. Particularly, as it later became clear by the work in \cite{Olshevsky2008}, it allowed studying switched linear systems for which there does not exist a common quadratic Lyapunov function.
	
	Along this line of thought, in this paper we aim to exploit nonlinear Perron-Frobenius theory \cite{LemmensNussbaum2012}, a generalization of non-negative matrix theory, to address nonlinear interactions in MASs without Lyapunov based arguments. It follows that a MAS modeled by a non-negative row-stochastic matrix is a particular case of the proposed general theory. The literature on nonlinear consensus problems is vast. It is mostly composed by particular nonlinear consensus protocols which offer advantages such as finite-time convergence \cite{Fran2015,Fran2017}, resilience to non-uniform time-delays \cite{Sun2009} and many more. These protocols are usually proved to converge to the consensus state via ad hoc Lyapunov functions. Most of approaches which aim to establish convergence to consensus for some class of nonlinear MAS falls in the general convexity theory of \cite{Moreau05}, i.e., each agent's next state is strictly inside the convex hull spanned by the state value of its neighbors. We mention the work in \cite{ZhiyunLin2007}, which is the continuous-time counterpart to the result of Moreau in \cite{Moreau05}, where the authors identify a class of non-linear interactions denoted as \emph{sub-tangent} and establish necessary and sufficient conditions on the network topology for convergence to consensus.
	
	Our approach sharply differs from the previous literature. We identify a class of functions (which comprises also non-negative row-stochastic matrices) which we prove to have a special convergence properties in the positive orthant $\mathbb{R}^n_{\geq 0}$. In particular, we take inspiration from nonlinear Perron-Frobenius theory and considered order-preserving and sub-homogeneous nonlinear maps. Furthermore, the approach presented in this paper differs significantly from the preliminary results presented in \cite{Deplano18} in both the statement of the theorems, lemmas, and their proof. 
	
	The \textbf{main contribution} of this paper is threefold. First, we provide sufficient conditions for stability of a class of nonlinear discrete-time systems represented by positive, sub-homogeneous, type-K order preserving maps. Second, we propose a sufficient condition on the structure of heterogeneous local interaction rules among agents which guarantees that the global model of the MAS falls into the considered class of nonlinear discrete-time systems. Third, we propose a sufficient condition which links the topology of the network and the structure of the local interaction rules to guarantee the achievement of a consensus state, i.e., the network state in which all state variables have the same value. Our results are a generalization to nonlinear discrete-time system of non-negative matrix theory applied to multi-agent systems, in so doing our results do not exploit Lyapunov function arguments. 

This paper is organized as follows. In Section \ref{sec:background} we present our notation and background material on multi-agent systems, order-preserving and sub-homogenoeus maps and recall the concept of periodic fixed-points. In Section \ref{sec:mainresults} we state our main results which consists in the statement of three main theorems. In Section \ref{shlsop} we discuss the proof of our results, we first discuss and list the required technical lemmas and then present the proof of each theorem in separate subsections. In Section \ref{sec:examples} we present examples of application of our theoretical results. Finally, in Section \ref{conclusion} we give our concluding remarks. 

\section{Background} \label{sec:background}
In this work we propose novel tools to perform stability analysis (consensus as a special case) of MASs whose state update is represented by \emph{positive}, \emph{order-preserving} and \emph{sub-homogeneous} maps. In this section we define a model of autonomous nonlinear MASs in discrete-time, its associated graph and present the above mentioned properties which define the class of MAS under study.

\subsection{Multi-agent systems}

We consider a MAS composed by a set of agents $V=\left\{1,\ldots,n\right\}$, which are modeled as autonomous discrete-time dynamical systems with scalar state in $\mathbb{R}_{\geq 0}=\{x\in\mathbb{R}:x\geq 0\}$. Agents are interconnected and update their state as follows
\begin{equation}\label{eq:locdiscretesystem}
\begin{array}{c}
x_{1}(k+1)=f_1\left(x_1(k),\ldots,x_n(k)\right)\\
\vdots\\
x_{n}(k+1)=f_n\left(x_1(k),\ldots,x_n(k)\right)
\end{array}\quad,
\end{equation} where $k\in\{0,1,2,\ldots\}$ is a discrete-time index. Introducing the
aggregate state $x =\left[x_{1},\ldots,x_{n}\right]^T\in \mathbb{R}^n$, system \eqref{eq:locdiscretesystem} can be written as
\begin{equation}\label{eq:globdiscretesystem}
	x(k+1) = f(x(k))
\end{equation} where $f:\mathbb{R}_{\geq 0}^n\rightarrow \mathbb{R}_{\geq 0}^n$ is differentiable. Hence, in this work we consider \emph{positive} systems \cite{Valcher18}. Positivity is a term with different meanings in different contexts, here by positive system we denote a system (and the associated map) with state that evolves in $\mathbb{R}_{\geq 0}^n$.

\begin{defn}[\textbf{Positive systems and maps}]\label{def:positivesystem}
~
System \eqref{eq:globdiscretesystem} is called positive if $f$ maps non-negative vectors into non-negative vectors, i.e., $f : \mathbb{R}_{\geq 0}^n \rightarrow \mathbb{R}_{\geq 0}^n$. Correspondingly, map $f$ is said to be positive. \hfill$\blacksquare$
\end{defn}

We now associate to the map $f$ a graph $\mathcal{G}(f)$ which captures the pattern of interactions among agents and denote it as \emph{inference graph} \cite{Liu13}.

Let $\mathcal{G}=(\mathcal{V},\mathcal{E})$ be a graph where $\mathcal{V}=\left\{1,\ldots, n\right\}$ is the set of nodes representing the agents and $\mathcal{E}\subseteq \mathcal{V} \times \mathcal{V}$ is a set of directed edges. A directed edge $(i,j)\in \mathcal{E}$ exists if node $i$ sends information to node $j$. To each agent $i$ is associated a set of nodes called neighbors of agent $i$ defined as $\mathcal{N}_i=\left\{j\in \mathcal{V}: (j,i)\in \mathcal{E}\right\}$. A \emph{directed path} between two nodes $p$ and $q$ in a graph is a finite sequence of $m$ edges $e_k=(i_k,j_k)\in E$ that joins node $p$ to node $q$, i.e., $i_1=p$, $j_m=q$ and $j_k=i_{k+1}$ for $k=1,\ldots,m-1$. A node $j$ is said to be \emph{reachable} from node $i$ if there exists a directed path from node $i$ to node $j$. A node is said to be \emph{globally reachable} if it is reachable from all nodes $i\in \mathcal{V}$.
 
\begin{defn}[\textbf{Inference graph}] \label{def:inferencegraph}
~
Given a map $f$ its inference graph $\mathcal{G}(f) = (\mathcal{V}, \mathcal{E})$ is defined by a set of nodes $\mathcal{V}$ and a set of directed edges $\mathcal{E}\subseteq \left\{\mathcal{V}\times \mathcal{V}\right\}$. An edge $(i,j)\in \mathcal{E}$ from node $i$ to node $j$ exists if 
\begin{equation*}
\frac{\partial f_i(x)}{\partial x_j}\neq 0\quad x \in \mathbb{R}_{\geq 0}^n \setminus S,
\end{equation*}
where $S$ is a set of measure zero in $\mathbb{R}^n$. \hfill $\blacksquare$
\end{defn}

\subsection{Order-preserving maps}\label{Intro:nonlinearPerron}

The set of $\mathbb{R}^n_{\geq 0}$ is a partially ordered set with respect to the natural order relation $\leq$. For $u,v\in \mathbb{R}^n_{\geq 0}$, we can write the partial ordering relations as follows
\begin{align*}
u\leq v &\Leftrightarrow u_i\leq v_i \quad \forall i\in \mathcal{V}, \\
u\lneq v &\Leftrightarrow u\leq v \text{ and } u\neq v,\\
u< v &\Leftrightarrow u_i< v_i \quad \forall i \in \mathcal{V}.
\end{align*}
The partial ordering $\leq$ yields an equivalence relation $\sim$ on $\mathbb{R}^n_{\geq 0}$, i.e., $x$ is equivalent to $y$ ($x \sim y$) if there exist $\alpha,\beta \geq 0$ such that $x\leq \alpha y$ and $y\leq \beta x$. The equivalence classes are called \emph{parts} of the cone of non-negative real vectors and the set of all parts is denoted by $\mathcal{P}$. It can be shown (see \cite{GaubertAkian2006}) that the cone $\mathbb{R}_{\geq 0}^n$ has exactly $2^n$ parts, which are given by $$P_I=\{x\in\mathbb{R}_{\geq 0}^n\:|x_i>0,\:\forall i\in I \text{ and } x_i=0 \text{ otherwise} \}\:,$$ with $I\subseteq \{1,\ldots,n\}$. We define a partial ordering on the set of parts $\mathcal{P}$ given by $P_{I_1} \preceq P_{I_2}$ if $I_1\subseteq I_2$. 

Maps which preserve such a vector order are said to be \emph{order-preserving}. Next, we provide a formal definition of three kinds of order-preserving maps present in the current literature.

\begin{defn}[\textbf{Order-preservation}]\label{def:ssop}
~
A positive map $f$ is said to be

\begin{itemize}
\item Order-preserving, if $\forall x,y \in \mathbb{R}^n_{\geq 0}$ it holds $$x\leq y \Leftrightarrow f(x)\leq f(y).$$
\item Strictly order-preserving, if if $\forall x,y \in \mathbb{R}^n_{\geq 0}$ it holds $$x\lneq y \Leftrightarrow f(x)\lneq f(y).$$
\item Strongly order-preserving, if $\forall x,y \in \mathbb{R}^n_{\geq 0}$ it holds
\begin{equation*}
x\lneq y \Leftrightarrow f(x)< f(y).\tag*{$\blacksquare$}
\end{equation*}
\end{itemize}

\end{defn}



The next remark is in order to clarify the context of the contribution of this paper. 

\begin{rem}
~
For linear maps, order-preservation and positivity are equivalent properties and correspond to non-negative matrices. Since this is not the case for general nonlinear maps, in this work we consider positive nonlinear maps which are also order-preserving.\hfill $\blacksquare$
\end{rem}


Now, we are ready to introduce the definition of \emph{type-K order-preserving} maps, shown next, which plays a pivotal role in the characterization of the class of nonlinear systems in which we are interested and which will be discussed at length in the proofs of our results. 

\begin{defn}[\textbf{Type-K Order-preservation}]\label{def:lsop}
~
A positive map $f$ is said to be type-K order-preserving if for any $x,y\in \mathbb{R}_{\geq 0}^n$ and $x\lneq y$ it holds 
\begin{enumerate}[label=$(\roman*)$]
\item $x_i = y_i \Rightarrow f_i(x)\leq f_i(y)$ , 
\item $x_i< y_i \Rightarrow f_i(x) < f_i(y)$ ,
\end{enumerate}
 for all $i=1,\ldots,n$, where $f_i$ is the $i$-th component of $f$.\hfill$\blacksquare$
\end{defn}

As it will be shown later, such a property is sufficient but not necessary for classical order-preservation. However, since it is easily identifiable from the sign structure of the Jacobian matrix, it allows to easily establish order-preservation of a given function. Furthermore, it constraints the behavior of the system, preventing the system from evolving with periodic trajectories and thus helping in proving convergence to a steady state.

\subsection{Sub-homogeneous maps}

Order-preserving dynamical systems and nonlinear Perron-Frobenius theory are closely related.
	In the theory of order-preserving dynamical systems, the emphasis is placed on strong order-preservation. For discrete-time strongly order-preserving dynamical systems one has generic convergence to periodic trajectories under appropriate conditions \cite{Polacik1992}. An extensive overview of these results was given by Hirsch and Smith \cite{Hirsch2006}.
	On the other hand, in nonlinear Perron-Frobenius theory one usually considers discrete-time dynamical systems that need not be strongly order-preserving, but satisfy an additional concave assumption and obtain similar results regarding periodic trajectories \cite{Lemmens2006nlp}. The concave assumption of interest in this paper is sub-homogeneity.

\begin{defn}[\textbf{Sub-homogeneity}]\label{def:sub-homogeneity}
~
A positive map is said to be sub-homogeneous if $$\alpha f(x) \leq f(\alpha x)$$ for all $x\in \mathbb{R}_{\geq 0}^n$ and $\alpha \in [0,1]$.\hfill $\blacksquare$
\end{defn}

Order-preserving and sub-homogeneous maps arise in a variety of applications, including optimal control and game theory \cite{GaubertAkian2003}, mathematical biology \cite{Rosenberg2001}, analysis of discrete event systems \cite{Gunawardena2003} and so on.

\subsection{Periodic points}

Concluding this section, we recall some basic concepts on periodic points which are instrumental to state our main results. 

Consider the state trajectory of the system in eq. \eqref{eq:globdiscretesystem}. A point $x\in \mathbb{R}^n$ is called a \emph{periodic point} of map $f$ if there exist an integer $p\geq 1$ such that $f^p(x)=x$. The minimal such $p\geq 1$ is called the \emph{period} of $x$ under $f$.

If $f(x)=x$, we call $x$ a fixed point of $f$. A \emph{fixed point} is a periodic point with period $p=1$. Fixed points of a map are equilibrium points for a dynamical system. We denote $F_f=\{x\in X:f(x)=x\}$ the set of all fixed points of map $f$.

The \emph{trajectory} of the system in eq. \eqref{eq:globdiscretesystem} with initial state $x$ is given by $\mathcal{T}(x,f)=\{f^k(x):k\in \mathbb{Z}\}$. 
If $f$ is clear from the context, we simply write $\mathcal{T}(x)$ to denote its trajectory, where $x$ is the initial state. If $x$ is a periodic point, we say that $\mathcal{T}(x)$ is a periodic trajectory. 
We denote the limit set of a point $x$ of map $f$ as $\omega(x,f)$ (or simply $\omega(x)$ if $f$ is clear from the context), which is defined as $$\omega(x)=\bigcap_{k\geq 0} cl\left(\{f^m(x):m\geq k\}\right),$$ with $cl(\cdot)$ denoting the closure of a set, i.e., the set together with all of its limit points. If $x$ is a fixed point it follows that the set $\omega(x)$ is a singleton, i.e., a set containing a single point.

\section{Main results} \label{sec:mainresults}

In this section we state and clarify the main results of this paper, while the following sections are dedicated to their proof.

For positive maps which are also order-preserving and sub-homogeneous, existing results (see next section for insights) do not provide any condition to ensure convergence to a fixed point, but only to periodic points \cite{Lemmens2006nlp} when the initial state is strictly positive, i.e., $x\in\mathbb{R}^n_+$. Furthermore, to the best of our knowledge, no result provides any information about trajectories whose initial state lies in the boundary of $\mathbb{R}_{\geq 0}^n$. 

Our aim is thus to fill this void considering the previously defined class of order-preserving maps, called \emph{type-K order-preserving}, for which we prove convergence to a fixed point for any initial state $x\in\mathbb{R}_{\geq 0}^n$ (and not only for $x\in\mathbb{R}^n_+$). This result is given in next theorem.

\begin{thm}[\textbf{Convergence}]\label{th:supnormembed}
~
Let a positive map $f$ be sub-homogeneous and type-K order-preserving. If $f$ has at least one positive fixed point in $\mathbb{R}^n_+$ then all periodic points are fixed points, i.e., the set $\omega(x)$ is a singleton and $\forall x\in \mathbb{R}_{\geq 0}^n: \lim_{k \rightarrow \infty} f^k(x)=\bar{x}$ where $\bar{x}$ is a fixed point of $f$.\hfill$\blacksquare$
\end{thm}


Using this technical result, another of our contributions is a sufficient condition on the heterogeneous local interaction rules under which a MAS is stable, i.e., its state converges to a fixed point. Fixed points are synonymous for equilibrium points, while in the literature the term fixed points is widely used in the context of iterated maps, the term equilibrium point is usually preferred in the context of discrete-time dynamical systems. This result is given in Theorem \ref{th:nonlinearconvergence}, whose statement is shown next.

\begin{thm}[\textbf{Stability}]\label{th:nonlinearconvergence}
~
Consider a MAS as in \eqref{eq:globdiscretesystem} with at least one positive equilibrium point. If the set of differentiable local interaction rules $f_i$, with $i= 1,\ldots,n$, satisfies the next conditions:

\begin{enumerate}[label=$(\roman*)$]
\item $f_i(x) \in \mathbb{R}_{\geq 0}$ for all $x\in \mathbb{R}_{\geq 0}^n$;
\item $\partial f_i/\partial x_i> 0$ and $\partial f_i/\partial x_j\geq 0$ for $i\neq j$;
\item $\alpha f_i(x) \leq f_i(\alpha x)$ for all $\alpha\in[0,1]$ and $x\in K$;
\end{enumerate}
then the MAS converges to one of its equilibrium points for any positive initial state $x(0)\in\mathbb{R}_{\geq 0}^n$.
\hfill$\blacksquare$
\end{thm}

As a special case, we also study the consensus problem for the considered class of MAS. We propose a sufficient condition based on the result in Theorem \ref{th:nonlinearconvergence} so that, for any initial state in $\mathbb{R}^n_+$, the MAS asymptotically reaches the consensus state, i.e., all state variable converge to same value. The proposed sufficient condition is graph theoretical and based on the inference graph $\mathcal{G}(f)$. The condition is satisfied if there exists a globally reachable node in graph $\mathcal{G}(f)$ and the consensus state is a fixed point for the considered MAS. This result is given in the next theorem.

\begin{thm}[\textbf{Consensus}]\label{th:nonlinearconsensus}
~
Consider a MAS as in \eqref{eq:globdiscretesystem}. If the set of differentiable local interaction rules $f_i$, with $i= 1,\ldots,n$, satisfies the next conditions:
\begin{enumerate}[label=$(\roman*)$]
\item $f_i(x)\in \mathbb{R}_{\geq 0}$ for all $x\in \mathbb{R}_{\geq 0}^n$;
\item $\partial f_i/\partial x_i> 0$ and $\partial f_i/\partial x_j\geq 0$ for $i\neq j$;
\item $\alpha f_i(x) \leq f_i(\alpha x)$ for all $\alpha\in[0,1]$ and $x\in \mathbb{R}^n_{\geq 0}$;
\item $f_i(x)=x_i$ if $x_i=x_j$ for all $j\in\mathcal{N}_{i}^{in}$;
\item Inference graph $\mathcal{G}(f)$ has a globally reachable node;
\end{enumerate}
then, the MAS converges asymptotically to a consensus state for any initial state $x(0)\in\mathbb{R}_{\geq 0}^n$.\hfill$\blacksquare$
\end{thm}

In the remainder of the paper, we discuss the proof of our main results in Theorem \ref{th:supnormembed}, \ref{th:nonlinearconvergence} and \ref{th:nonlinearconsensus}. 

\section{Proof of main results} \label{shlsop}

We begin by clarifying the relationships among the different kinds of order-preservation.

\begin{rem}\label{rem:lorderpreservation}
~
Strong order-preservation $\Rightarrow$ Type-K order-preservation $\Rightarrow$ strict order-preservation $\Rightarrow$ order-preservation.\hfill$\blacksquare$
\end{rem}

Every converse relationship in Remark \ref{rem:lorderpreservation} does not hold. Given $x,y\in\mathbb{R}$, let $f:\mathbb{R}^2\rightarrow\mathbb{R}^2$, we have the following counter-examples:
\begin{itemize}
\item $f(x,y)=[1,1]^T$ is order-preserving but not strictly;
\item $f(x,y)=[y,x]^T$ is strictly order-preserving but not type-K;
\item $f(x,y)=[\sqrt{x}+y,y]^T$ is type-K order-preserving but not strongly.
\end{itemize}

Usually, to verify order-preservation is not an easy task. For differentiable continuous-time systems $\dot{x}=f(x)$ a sufficient condition to ensure order-preservation is given by Kamke \cite{Smith88,Kamke32}. The Kamke condition usually exploited in the analysis of continuous time systems is shown next.

\begin{lem}[\textbf{Kamke Condition}]\label{th:ifflsop}
~\cite{Smith88,Kamke32}~
The map $f$ of a continuous-time system $$\dot{x}=f(x)$$ is order-preserving if its Jacobian matrix is Metzler, i.e.,
\begin{equation*}
\partial f_i/\partial x_j\geq 0\text{ for }i\neq j\:.\tag*{$\blacksquare$}
\end{equation*}
\end{lem}

As a counterpart to Lemma \ref{th:ifflsop}, for discrete-time systems we propose a sufficient condition to ensure type-K order-preservation of a map $f$, instrumental to the analysis of discrete-time systems, which we denote \emph{Kamke-like} condition.

\begin{prop}[\textbf{Kamke-like condition}]\label{th:ifflsoplike}
~
The map $f$ of a discrete-time system $$x(k+1)=f(x(k))$$ is type-K order-preserving if its Jacobian matrix is Metzler with strictly positive diagonal elements, i.e., if 
\begin{equation} \label{condition}
\partial f_i/\partial x_i> 0\text{ and } \partial f_i/\partial x_j\geq 0\text{ for }i\neq j\:.
\end{equation}
\end{prop}

\begin{pf*}{\textbf{Proof}.}
Let $x\in \mathbb{R}^n$ and, without lack of generality, $y=x+\varepsilon e_j$ where $\varepsilon >0$ and $e_j$ denotes a canonical vector with all zero values but the $j$-th which is $1$, thus $x\lneq y$. If \eqref{condition} holds, then
\begin{enumerate}
\item If $i\neq j$ then $y_i=x_i+\varepsilon 0=x_i$ and $$\frac{\partial f_i(x)}{\partial x_j}= \lim_{\varepsilon\rightarrow 0} \frac{f_i(x+\varepsilon e_j)-f_i(x)}{\varepsilon}\geq0\:,$$ which implies that $f_i(x)\leq f_i(x+\varepsilon e_j)=f_i(y)$, i.e., condition $(i)$ of Definition \ref{def:lsop}.
\item If $i = j$ then $y_i=x_i+\varepsilon 1>x_i$ and $$\frac{\partial f_i(x)}{\partial x_i}= \lim_{\varepsilon\rightarrow 0} \frac{f_i(x+\varepsilon e_i)-f_i(x)}{\varepsilon}>0\:,$$ which implies that $f_i(x)<f_i(x+\varepsilon e_i)=f_i(y)$, i.e., condition $(ii)$ of Definition \ref{def:lsop}.
\end{enumerate}
Since $1)\Rightarrow \eqref{condition}$ and $2) \Rightarrow \eqref{condition}$, the proof is complete.\hfill $\square$
\end{pf*}

Having clarified how to verify the type-K order-preserving property for a discrete-time system, we move on in the next subsection to discuss a significant property of order-preserving maps which are also sub-homogeneous, i.e., non-expansivness with respect to the so-called Thompson's metric.

\subsection{Non-expansive maps}

Dynamical systems defined by order-preserving and sub-homogeneous maps are non-expansive under the Thompson's metric. Here, we introduce the concepts of \emph{non-expansiveness} and \emph{Thompson's metric} and give a few useful lemmas.

\begin{defn}[\textbf{Thompson's metric} \cite{Thompson64}]\label{def:thompson}
~
For $x,y\in\mathbb{R}^n_{\geq 0}$ define $$M(x/y)=\inf\{\alpha\geq 0: y\leq \alpha x\}\:,$$ with $M(x/y)=\infty$ if the set is empty. By mean of function $M(y/x)$, Thompson's metric $d_T:\mathbb{R}^n\times\mathbb{R}^n\rightarrow[0,\infty]$ is defined for all $(x, y)\in (\mathbb{R}^n\times \mathbb{R}^n)\setminus (0,0)$ as follows $$d_T(x,y)=\log(\max\{M(x/y),M(y/x)\})$$ with $d_T(0, 0) = 0$. \hfill $\blacksquare$
\end{defn}

\begin{defn}[\textbf{Non-expansiveness}]\label{def:non-expansiveness}
~
A positive map $f$ is called non-expansive with respect to a metric $d:\mathbb{R}^n\times\mathbb{R}^n\rightarrow \mathbb{R}_{\geq 0}$, if $$d(f(x),f(y))\leq d(x,y)$$ for all $x,y\in\mathbb{R}_{\geq 0}^n$. \hfill $\blacksquare$
\end{defn}

The next result taken from \cite{GaubertAkian2006} but stated in reference to positive cones $K$. In this paper we always consider as a particular case the cone of non-negative vectors, i.e., $K=\mathbb{R}^n_{\geq0}$, which is a solid, closed and convex cone.

\begin{lem} \label{lem:non-expansiveness} 
~\cite{GaubertAkian2006}~
Let $K$ be a solid closed convex cone\footnote{A set $K\in\mathbb{R}^n$ is called a \emph{convex cone} if $\alpha K \subseteq K$ for all $\alpha \geq 0$ and $K \cap (-K) = \{0\}$. The convex cone $K$ is \emph{closed} if it is a closed set in $\mathbb{R}^n$ and it is solid if it has a non-empty interior.} in $\mathbb{R}^n$. If $f:K\rightarrow K$ is an order-preserving map, then it is sub-homogeneous if and only if it is non-expansive with respect to Thompson's metric $d_T$. \hfill $\blacksquare$
\end{lem}

Such a property allows one to prove detailed results concerning the behavior of dynamical systems, see \cite{GaubertAkian2006,Lemmens2005,Nussbaum1990} and also Chapter~8 in \cite{LemmensNussbaum2012} and reference therein. 

Thompson's metric $d_T$ and the sup-norm $\left\lVert{\cdot}\right\rVert_{\infty}$ defined by $$\left\lVert{x}\right\rVert_{\infty}=\max_i \left\|x_i\right\|\:,$$ are closely related thanks to the following lemma.

\begin{lem} \label{lem:isometry} 
~\cite{Thompson64}~
The coordinate-wise logarithmic function $L:\mathbb{R}^n_{+}\rightarrow\mathbb{R}^n$ is an isometry from $(\mathbb{R}^n_{+},d_T)$ to $(\mathbb{R}^n,\left\lVert\cdot\right\rVert_{\infty})$, with $\mathbb{R}_+=\{x\in \mathbb{R}:x>0\}$.\hfill$\blacksquare$ 
\end{lem}

By Lemma \ref{lem:isometry}, if a positive map $f$ can be restricted to $\mathbb{R}_+^n=int(\mathbb{R}_{\geq 0}^n)$ and if it is order preserving and sub-homogenous, then $g:\mathbb{R}^n\rightarrow\mathbb{R}^n$ given by $g = log \circ f \circ exp$, is a sup-norm non-expansive map that has the same dynamical properties as $f$. The dynamics of sup-norm non-expansive maps is widely known. In fact, there exists the following result, which is the simplified version of Theorem 4.2.1 in \cite{LemmensNussbaum2012}.

\begin{lem}\label{lem:boundedness} 
~\cite{Thompson64}~
If $f:\mathbb{R}^n\rightarrow\mathbb{R}^n$ is a sup-norm non-expansive then only one of the next cases can occur:
\begin{enumerate}[label=$(\roman*)$]
\item $\forall x\in\mathbb{R}^n$ trajectories $\mathcal{T}(x)$ are unbounded;
\item $\forall x\in\mathbb{R}^n$ trajectories $\mathcal{T}(x)$ are bounded.\hfill $\blacksquare$
\end{enumerate} 
\end{lem}


\subsection{Proof of Theorem \ref{th:supnormembed}}

As pointed out in the previous section, for positive maps $f$ which are also order-preserving and sub-homogeneous it is possible to establish the boundedness of any trajectory with initial states $x(0)\in \mathbb{R}_+^n$ and entirely enclosed in $\mathbb{R}_+^n$. On the contrary, there are still no results for trajectories with initial state $x(0)$ in the boundary of $\mathbb{R}_{\geq 0}^n$.

When $f$ satisfies the additional property of type-K order preservation, Theorem \ref{th:supnormembed} gives a sufficient condition for the boundedness of any trajectory with initial state $x(0)\in\mathbb{R}_{\geq 0}^n$ and convergence of such trajectories to a fixed point of map $f$.

The proof of Theorem \ref{th:supnormembed} requires the preliminary discussion of several lemmas which are needed to prove that type-K order-preservation as opposed to simple order-preservation, together with other properties, is sufficient to extend result to trajectories starting at any point in $\mathbb{R}^n_{\geq 0}$ and exclude the existence of periodic trajectories.

\begin{lem}\label{lem:zeropattern}
~
Let a positive map $f$ be type-K order-preserving. For all $x\in\mathbb{R}_{\geq 0}^n$ it holds that $f_i^k(x)>0$ for all $i$ such that $x_i>0$ and $k\geq 1$.
\end{lem}
\begin{pf*}{\textbf{Proof}.}
For any $x \in\mathbb{R}_{\geq 0}^n$ let $I(x)\subset\{1,\ldots, n\}$ be such that $x_i>0$ for $i\in I(x)$ and $x_i=0$ otherwise. Since $\mathbf{0}\leq x$, by type-K order-preservation of $f$ follows $f(\mathbf{0})\leq f(x)$. More precisely it holds $f_i(x)> f_i(\mathbf{0}) \geq 0$ for $i\in I(x)$ and $f_i(x)\geq f_i(\mathbf{0}) \geq 0$ otherwise, implying $I(x)\subseteq I(f(x))$. By induction, $I(x)\subseteq I(f^k(x))$, i.e., $f^k_i(x)>0$ for all $i\in I(x)$, completing the proof. \hfill $\square$
\end{pf*}

\begin{lem}\label{lem:partsordering}
~
Let a positive map $f$ be sub-homogeneous and type-K order-preserving. For all $x\in\mathbb{R}_{\geq 0}^n$ there exists a part $P\in\mathcal{P}(\mathbb{R}_{\geq 0}^n)$ and an integer $k_0\in \mathbb{Z}$ such that $f^{k}(x)\in P$ for all $k\geq k_0$. 
\end{lem}
\begin{pf*}{\textbf{Proof}.}
Since by Lemma \ref{lem:non-expansiveness} $f$ is non-expansive under the Thompson's metric $d_T$, then $x\sim y$ implies $f(x)\sim f(y)$. This can be easily proved by noticing that $d_T(f(x),f(y))\leq d_T(x,y)<\infty$ since $x\sim y$.
This means that $f$ maps parts into parts, i.e., for all $x\in\mathbb{R}_{\geq 0}^n$ and $x'\in [x]=P_{I_0}$ it holds $f(x')\in[f(x)] = P_{I_1}$. By Lemma \ref{lem:zeropattern} it follows $P_{I_0}\preceq P_{I_1}$ and therefore $[x]\preceq [f(x)]$.	
Generalizing, we say that $f^k(x)\in P_{I_k}$ with $k\in \mathbb{Z}$ and $I_{k}\subseteq I_{k+1}\subseteq\{1,\ldots,n\}$. There exists $k_0\in \mathbb{Z}$ such that $I_k=I_{k_0}$ for all $k> k_0$ and thus $P_k=P_{k_0}$. This completes the proof.\hfill $\square$
\end{pf*}

\begin{lem}\label{lem:boundedTrajectory}
~
Let a positive map $f$ be sub-homogeneous and type-K order-preserving. If $f$ has a positive fixed point $\bar{x}\in\mathbb{R}_+^n$, then for all $x\in\mathbb{R}_{\geq 0}^n$ the trajectory $\mathcal{T}(x)$ is bounded.
\end{lem}
\begin{pf*}{\textbf{Proof}.}
By Lemma \ref{lem:isometry} it follows that $g=\log\circ f \circ \exp$ is a sup-norm non-expansive map that has the same dynamical properties as $f$ for all $x\in \mathbb{R}_+^n$. By Lemma \ref{lem:boundedness} we know that one of the two cases can occur:

\begin{enumerate}[label=$(\roman*)$]
\item all trajectories $\mathcal{T}(\log(x),g)$ are unbounded;
\item all trajectories $\mathcal{T}(\log(x),g)$ are bounded.
\end{enumerate}

Since $f$ has a fixed point $x_f\in \mathbb{R}_+^n$, such that $f(x_f)=x_f$, then $x_g=\log(x_f)$ is a fixed point of $g$, i.e., $g(x_g)=x_g$. The trajectory $\mathcal{T}(\log(x_f),g)$ is obviously bounded and therefore case $(ii)$ holds. 

By Lemma \ref{lem:partsordering}, we can partition $\mathbb{R}_{\geq 0}^n$ in two disjoint sets $S_1$, $S_2$ such that if for $x$ there exists $k_0\in\mathbb{Z}$ such that $f^{k_0}(x) \in \mathbb{R}_+^n$, then $x\in S_1$, otherwise $x\in S_2$. We analyze these two cases.

1) For all $x\in S_1$, by Lemma \ref{lem:partsordering}, it holds that $f^{k}(x) \in \mathbb{R}_+^n$ for all $k\geq k_0$. Let $x_0 = f^{k_0}(x)$. Since case $(ii)$ holds $\mathcal{T}(\log(x_0),g)$ is bounded, because of the isometry also $\mathcal{T}(x_0,f)$ is bounded, and therefore also $\mathcal{T}(x,f)$. We conclude that for all $x\in S_1$ trajectories $\mathcal{T}(x,f)$ are bounded.
	
2) For all $x\in S_2$, by Lemma \ref{lem:partsordering}, there exists $k_0\in\mathbb{Z}$ such that $f^k(x)\in P_I$ with $I(x)\subset N = \{1,\ldots,n\}$ for all $k\geq k_0$.
	Without loss of generality, here we assume $I=\{1,\ldots,m\}$, where $m<n$.
	Let $x =[z_1^T,z_2^T]^T$ with $z_1\in \mathbb{R}^m_{\geq 0}$ and consider the following $m$-dimensional map $f^*:\mathbb{R}_+^m\rightarrow \mathbb{R}_+^m$ defined by $$f^*_i(z_1)=f_i(z_1,z_2),\quad z_2=\mathbf{0}\:,$$ with $i\in I(x)$.
	It is not difficult to check that $f^*$ is still sub-homogeneous and type-K order preserving. Accordingly, $g^*=\log\circ f^* \circ \exp$ is a sup-norm non-expansive map that has the same dynamical properties as $f^*$ for all $x\in \mathbb{R}_+^m$.
	The main point now is to prove that if $(ii)$ occurs then all trajectories $\mathcal{T}(\log(z_1),g^*)$ are also bounded. To this aim, we first need to show that for all $i\in I(x)$ it holds 
	\begin{equation}\label{eq:gis}
	g^*_i(z_1)\leq g_i(z_1,z_2).
	\end{equation} 
	Since both the exponential and the logarithmic functions are strictly increasing, \eqref{eq:gis} is equivalent to 
	\begin{equation}\label{eq:fis}
	f^*_i(z_1)\leq f_i(z_1,z_2).
	\end{equation}
		By definition, \eqref{eq:fis} holds if $z_2 = 0$. If $z_2\neq 0$, for any $x=[z_1^T,z_2^T]^T$ consider $\bar{x}=[z_1^T,\bar{z}_2^T]^T$ such that $\bar{z}_2=\mathbf{0}$. Since $f$ is order-preserving, for all $i\in I$ it holds that $f_i(\bar{x})\leq f_i(x)$, which is equivalent to write $f_i(z_1,\bar{z}_2)\leq f_i(z_1,z_2)$ . By definition, $f_i^*(z_1)=f_i(z_1,\bar{z}_2)$. Therefore, $f_i^*(z_1) \leq f_i(z_1,z_2)$ for all $z_2\neq 0$, i.e., \eqref{eq:fis} and \eqref{eq:gis} hold. 
	Suppose that $(ii)$ occurs and there exist $\hat{z}_1\in \mathbb{R}_+^m$ such that $\mathcal{T}(\log(\hat{z}_1),g^*)$ is unbounded. By \eqref{eq:gis} it is clear that given $\hat{x}=[\hat{z}_1^T,z_2^T]^T$ the trajectory $\mathcal{T}(\log(\hat{x}),g)$ is also unbounded, contradicting $(ii)$.
	Let $x_0 = f^{k_0}(x)$. Since all trajectories $\mathcal{T}(\log(x_0),g)$ are bounded, because of the isometry also $\mathcal{T}(x_0,f)$ is bounded, and therefore also $\mathcal{T}(x,f)$. We conclude that for all $x\in S_2$ trajectories $\mathcal{T}(x,f)$ are bounded.\hfill $\square$
\end{pf*}

\begin{lem}\label{lem:intsingleton}
~\cite{Jiang1996}~
Let a positive map $f$ be sub-homogeneous and type-K order-preserving. If for all $x\in K$ the trajectory $\mathcal{T}(x)$ is bounded, then for all $x\in \mathbb{R}_{\geq 0}^n$, $\omega(x)$ is a singleton.\hfill$\blacksquare$
\end{lem}

Finally, after restating the Theorem \ref{th:supnormembed} for convenience of the reader, we present a compact proof based on the results presented in this section and Lemma 3.1.3 in \cite{LemmensNussbaum2012}.

\textbf{Theorem 7 (Convergence)} $\text{  }$ \emph{Let a positive map $f$ be sub-homogeneous and type-K order-preserving. If $f$ has at least one positive fixed point in $\mathbb{R}^n_+$ then all periodic points are fixed points, i.e., the set $\omega(x)$ is a singleton and $\forall x\in \mathbb{R}_{\geq 0}^n: \lim_{k \rightarrow \infty} f^k(x)=\bar{x}$ where $\bar{x}$ is a fixed point of $f$}. 

\begin{pf*}{\textbf{Proof of Theorem \ref{th:supnormembed}}.}
~
By Lemma~\ref{lem:boundedTrajectory} it follows that all trajectories $\mathcal{T}(x)$ are bounded for all $x\in\mathbb{R}_+^n$. By Lemma \ref{lem:intsingleton} it follows that all periodic points are fixed points, i.e., the set $\omega(x)$ is a singleton. By Lemma 3.1.3 in \cite{LemmensNussbaum2012}, since $f$ is continuous all trajectories are bounded and all periodic points are fixed points it holds $\lim_{k \rightarrow \infty} f^k(x)=\bar{x}$ where $\bar{x}$ is a fixed point of $f$.\hfill $\square$
\end{pf*}


\subsection{Proof of Theorem \ref{th:nonlinearconvergence}}

By means of the technical result in Theorem \ref{th:supnormembed} we prove our second main result, a sufficient condition on the structure of the heterogeneous local interaction rules of the MAS under consideration so that the global map (possibly unknown due to an unknown network topology) is positive, type-K order preserving and sub-homogeneous map, thus falling within the class of systems considered in Theorem \ref{th:ifflsoplike}. We also restate the theorem for convenience of the reader. 

\textbf{Theorem 8 (Stability)} $\text{  }$ \emph{Consider a MAS as in \eqref{eq:globdiscretesystem} with at least one positive equilibrium point. If the set of differentiable local interaction rules $f_i$, with $i= 1,\ldots,n$, satisfies the next conditions:
\begin{enumerate}[label=$(\roman*)$]
\item $f_i(x) \in \mathbb{R}_{\geq 0}$ for all $x\in \mathbb{R}_{\geq 0}^n$;
\item $\partial f_i/\partial x_i> 0$ and $\partial f_i/\partial x_j\geq 0$ for $i\neq j$;
\item $\alpha f_i(x) \leq f_i(\alpha x)$ for all $\alpha\in[0,1]$ and $x\in K$;
\end{enumerate}
then the MAS converges to one of its equilibrium points for any positive initial state $x(0)\in\mathbb{R}_{\geq 0}^n$.}

\begin{pf*}{Proof of Theorem \ref{th:nonlinearconvergence}.}
We start the proof by establishing equivalence relationships between the properties $(i)-(iii)$ of the local interaction rules of the MAS listed in the statement of Theorem \ref{th:nonlinearconvergence} and properties (a)-(c) shown next:
\begin{enumerate}[label=$(\alph*)$]
\item $f$ is positive;
\item $f$ is type-K order-preserving;
\item $f$ is sub-homogeneous;
\end{enumerate}
We now prove all equivalences one by one.

\begin{itemize}

\item $[(i)\Leftrightarrow (a)]$ Condition $(i)$ implies that $f$ maps a point of $\mathbb{R}_{\geq 0}^n$ into $\mathbb{R}_{\geq 0}^n$ and is, therefore, a positive map (see Definition \ref{def:positivesystem}).

\item $[(ii)\Rightarrow (b)]$ due to Proposition \ref{th:ifflsoplike} (Kamke-like condition).

\item $[(iii)\Leftrightarrow (c)]$ by Definition \ref{def:sub-homogeneity} of a sub-homogeneous map, sub-homogeneity can be verified element-wise for map $f$, thus the equivalence follows.

\end{itemize}

Thus, if conditions $(i)$ to $(iii)$ hold true for all local interaction rules $f_i$ with $i=1,\ldots,n$, since by assumption map $f$ has at least one positive fixed point, we can exploit the result in Theorem \ref{th:supnormembed} to establish that for all positive initial conditions, the state trajectories of the MAS converge one of its positive equilibrium points. \hfill $\square$
\end{pf*}

\subsection{Proof of Theorem \ref{th:nonlinearconsensus}}

To prove our third main result, namely Theorem \ref{th:nonlinearconsensus}, we need to introduce two technical lemmas. The first lemma, shown next, states sufficient conditions under which the elements along the rows of the Jacobian matrix of a map $f$ computed at a consensus point $c \mathbf{1}$ sum to one.

\begin{lem}\label{lem:stochasticJ}
~
Let a map $f$ be positive and differentiable. If the set of fixed points $F_f$ of map $f$ satisfies $F_f\supseteq\{c\mathbf{1},\quad c\in\mathbb{R}_{\geq 0}\}$, i.e., the set of fixed points contains at least all positive consensus states, then $$J_f(c\mathbf{1})\mathbf{1}=\mathbf{1}\quad \forall c \in \mathbb{R}_+\:,$$ where $J_f(c\mathbf{1})$ denotes 
the Jacobian of $f$ evaluated in $c\mathbf{1}$.
\end{lem}
\begin{pf*}{\textbf{Proof}.} Since $f$ is differentiable, we can apply directly the definition of directional derivative in a point $x\in \mathbb{R}^n_{\geq 0}$ along a vector $v\in\mathbb{R}^n$ obtaining $$J_f(x)v = \lim_{h\rightarrow 0}\frac{f(x+hv)-f(x)}{h}\:.$$ Now we evaluate this expression in a consensus point $x=c\mathbf{1}\in F_f$, and along the direction $v=\mathbf{1}$ which is an invariant direction of $f$. We obtain 
\begin{align}
J_f(c\mathbf{1})\mathbf{1} &= \lim_{h\rightarrow 0}\frac{f(c\mathbf{1}+h\mathbf{1})-f(c\mathbf{1})}{h}\:, \nonumber \\
 &= \lim_{h\rightarrow 0}\frac{\cancel{c\mathbf{1}}+h\mathbf{1}-\cancel{c\mathbf{1}}}{h}= \mathbf{1}\:,\nonumber 
\end{align}
thus proving the statement. \hfill $\square$
\end{pf*}

Next, we introduce a critical lemma needed to prove our third main result. In particular, we show that if there exists a fixed point of map $f$ different from a consensus point, then there exists a consensus point such that the Jacobian of map $f$ computed at that consensus point has a unitary eigenvalue with multiplicity strictly greater than one. 
\begin{lem} \label{lem:invdir} 
~Let $f$ be positive, sub-homogeneous, type-K order-preserving and have a set of fixed points $F_f$ such that $$F_f\supseteq\{c\mathbf{1}, \ \ c\in \mathbb{R}_{\geq 0}\}\:.$$ 

If there exists a fixed point $\bar{x}\in\mathbb{R}^n_{\geq 0}$ such that $$\bar{x}\neq c\mathbf{1}\:,\quad \forall c\in\mathbb{R}_{\geq 0}$$ then there exists $\bar{c}(\bar{x})>0$ such that the Jacobian matrix $J_f(\bar{c}(\bar{x})\mathbf{1})$ of map $f$ computed at $\bar{c}(\bar{x})\mathbf{1}$ has a unitary eigenvalue with multiplicity strictly greater than one.

\end{lem}
\begin{pf*}{\textbf{Proof}.}
Let $\bar{x}=\left[\bar{x}_1,\ldots,\bar{x}_n\right]^T\in\mathbb{R}^n_{\geq 0}$ be a fixed point of map $f$ and let $c_1,c_2\in\mathbb{R}_{\geq 0}$ be such that
\begin{align*}
c_1 & = \min_{i=1,\ldots, n} \bar{x}_i\:,\\
c_2 & = \max_{i=1,\ldots, n} \bar{x}_i\:.
\end{align*}
We define three sets
\begin{align*}
I_{min}(\bar{x}) & =\{i:\bar{x}_i = c_1\},\\
I_{max}(\bar{x}) & =\{i:\bar{x}_i = c_2\},\\
I(\bar{x}) & =\{i:\bar{x}_i\neq c_1,c_2\}.
\end{align*}
Consider a point $y$ such that the $i$-th component is defined by
\begin{equation}\label{defy}
y_i=
\begin{cases} 
c_1 & \text{if } i\in I_{min}(\bar{x}) \\
c_3 & \text{otherwise}
\end{cases}
\end{equation}
and such that 
\begin{gather}\label{eq:res1}
c_1\mathbf{1}\lneq y \lneq \bar{x}\lneq c_2\mathbf{1}\:.
\end{gather}
By \eqref{defy} and \eqref{eq:res1} it follows that 
\begin{equation}\label{eq:res2}
y\leq c_3\mathbf{1}.
\end{equation}
$\bullet$ Now, we prove that
\begin{equation}\label{eq:ordflow}
f(y)\leq y.
\end{equation} 
Since map $f$ is type-K order preserving, from \eqref{eq:res1} it follows $c_1\leq f_i(y) \leq \bar{x}_i$ and from \eqref{eq:res2} $f_i(y) \leq c_3$ for $i=1,\ldots,n$. For $i \in I_{min}(\bar{x})$, by definition $\bar{x}_i=c_1$ and thus $f_i(y)=c_1$, otherwise for $i \in I(\bar{x})\cup I_{max}(\bar{x})$ by \eqref{eq:res1} $\bar{x}_i\geq y_i = c_3$ and it follows $c_1\leq f_i(y)\leq c_3$. Thus, \eqref{eq:ordflow} holds.

$\bullet$ Now, we prove that 
\begin{equation}\label{eq:constr2}
\frac{c_3}{c_2}\bar{x}_i\leq f_i(y)\leq c_3,\quad i=1,\ldots,n\:.
\end{equation}

Since $f$ is order-preserving and sub-homogeneous, then $f$ is non-expansive under the Thompson's metric (see Definition \ref{def:thompson}) by Lemma \ref{lem:non-expansiveness}. Now, by exploiting the definition of non-expansive map, we compute an upper bound to $d_T(\bar{x},f(y))$. It holds
\begin{align}
d_T(\bar{x},f(y)) &\leq d_T(\bar{x},y) \nonumber\\
 &=\log\left(\max \left\{ M(\bar{x}/y), M(y/\bar{x}) \right\}\right) \nonumber
\end{align}
where
\begin{align}
M(\bar{x}/y)= &\inf\{\alpha\geq 0: y\leq \alpha \bar{x}\}= \max_i \frac{y_i}{x_i}=1, \nonumber\\
M(y/\bar{x})= &\inf\{\alpha\geq 0: \bar{x}\leq \alpha y\} = \max_i \frac{x_i}{y_i}\leq \frac{c_2}{c_3}. \nonumber
\end{align}
Since $c_2\geq c_3$, it holds
\begin{equation}\label{upper}
d_T(\bar{x},f(y))\leq \frac{c_2}{c_3}.
\end{equation}
Now, we compute a lower bound to $d_T(\bar{x},f(y))$, where
\begin{align}
d_T(\bar{x},f(y))=\log\left(\max \left\{ M(\bar{x}/f(y)), M(f(y)/\bar{x}) \right\}\right) \nonumber
\end{align}
and
\begin{align}
M(\bar{x}/f(y))= &\inf\{\alpha\geq 0: f(y)\leq \alpha \bar{x}\}= \max_i \frac{f_i(y)}{x_i}=1, \nonumber\\
M(f(y)/\bar{x})= &\inf\{\alpha\geq 0: \bar{x}\leq \alpha f(y)\} = \max_i \frac{x_i}{f_i(y)}\geq \frac{c_2}{c_3}. \nonumber
\end{align}
Thus
$\max \left\{ M(\bar{x}/f(y)), M(f(y)/\bar{x}) \right\} \geq \frac{c_2}{c_3},$
therefore
\begin{equation}\label{lower}
d_T(\bar{x},f(y))\geq \frac{c_2}{c_3}.
\end{equation}
By inequalities \eqref{upper} and \eqref{lower} it follows that $$d_T(\bar{x},f(y))= \max_i \frac{x_i}{f_i(y)}= \frac{c_2}{c_3},$$ thus proving that the inequality in \eqref{eq:constr2} holds true.

$\bullet$ Due to Theorem \ref{th:supnormembed} it holds $\displaystyle \lim_{k\rightarrow\infty} f^k_i(y)=\bar{y}_i$. Now, we prove that 
\begin{equation} \label{13}
\bar{y}_i=
\begin{cases}
c_1 & \text{if } i\in I_{min}(\bar{x}),\\
c_3 & \text{if } i\in I_{max}(\bar{x}),\\
c_1 & \text{if } i\in I(\bar{x})\text{ and }\\ &\exists k^*:f^{k^*}_i(y)<f_i^{k^*-1}(y),\\
c_3 & \text{otherwise}.
\end{cases}
\end{equation}

In \eqref{13} three cases may occur:

\begin{enumerate}
\item If $i \in I_{min}(\bar{x})$ then $\bar{x}_i=c_1$ and by \eqref{eq:res1} it follows $f_i(y)=c_1$.

\item If $i \in I_{max}(\bar{x})$ then $\bar{x}_i=c_2$ and by \eqref{eq:constr2} it follows $f_i(y)=c_3$.

\item If $i \in I(\bar{x})$, by \eqref{eq:ordflow} two cases may occur:
\begin{enumerate}
\item There exists $k^*>0$ such that $f^{k^*}_i(y)<y_i$. In this case, by type-K order-preservation it holds that $f^k_i(y)<f_i^{k-1}(y)\quad \forall k\geq k^*+1$ and therefore $$\lim_{k\rightarrow \infty} f_i^k(y)=c_1.$$
\item Otherwise $f^k_i(y)=f^{k-1}(y)\quad \forall k>0$ and therefore $$\lim_{k\rightarrow \infty} f_i^k(y)=y_i=c_3.$$
\end{enumerate}
\end{enumerate}
Thus, by \eqref{13} we proved that for any fixed point $\bar{x}$ different from a consensus point $c\mathbf{1}$ there exists a fixed point $\bar{y}$ with elements corresponding to either $c_1$ or $c_3$ and such that $I(\bar{y})=\emptyset$.

Now, consider a point $z$ such that its $i$-th component is defined as follows
\begin{equation}\label{eq:newpoint}
z_i=
\begin{cases}
c_1 & \text{if } i \in I_{min}(\bar{y}) \\
c_4 & \text{if } i \in I_{max}(\bar{y})
\end{cases}
\end{equation}
with $c_4 \in \left[c_1,c_3\right]$.
By \eqref{13} and \eqref{eq:newpoint}, we can conclude that $z$ is fixed point, i.e., $f(z)=z$, for all values of $c_4$ in the interval $c_4\in \left[c_1, c_3\right]$.
Now, let $v(\bar{x})$ be a vector such that
\begin{equation}\label{defV}
v_i(\bar{x}) =
\begin{cases}
0 & \text{if } i\in I_{min}(\bar{x})\\
1 & \text{if } i\in I_{max}(\bar{x})\\
0 \; or \; 1 & \text{if } i\in I(\bar{x})
\end{cases}\:,
\end{equation}
Thus, by \eqref{defV} the point $c_1\mathbf{1}+hv(\bar{x})$ is a fixed point of map $f$ for all $h\in[0, c_3-c_1]$. Thus, it follows that $$f(c_1\mathbf{1}+hv(\bar{x})) = c_1\mathbf{1}+hv\:,\quad h\in[0, c_3-c_1]\:.$$ 

Since $v(\bar{x}) \neq \mathbf{1}$, it holds (by reasoning along the lines of Lemma \ref{lem:stochasticJ}) that the Jacobian of map $f$ computed at $c_1\mathbf{1}$ has a right eigenvector equal to $v(\bar{x})$, i.e., $J_f(c_1\mathbf{1})v(\bar{x})=v(\bar{x})$. By Lemma \ref{lem:stochasticJ} it holds that the Jacobian of $f$ satisfies $J_f(c\mathbf{1})\mathbf{1}=\mathbf{1}$ for all $c>0$. Thus, if there exists $\bar{x}\neq c \mathbf{1}$ then there exists $\bar{c}(\bar{x})=\displaystyle \min_{i=1,\ldots,n} \bar{x}_i=c_1$ such that matrix $J_f(\bar{c}(\bar{x})\mathbf{1})$ has a untiray eigenvalue with multiplicity strictly greater than one, thus proving the statement of this lemma. \hfill $\square$
\end{pf*}

Finally, we recall for convenience of the reader and detail a proof of our third and last main result, i.e., Theorem \ref{th:nonlinearconsensus}, which provides sufficient conditions for asymptotic convergence to the consensus state.

\textbf{Theorem 9 (Consensus)} $\text{  }$ \emph{Consider a MAS as in \eqref{eq:globdiscretesystem}. If the set of differentiable local interaction rules $f_i$, with $i= 1,\ldots,n$, satisfies the next conditions:
\begin{enumerate}[label=$(\roman*)$]
\item $f_i(x)\in \mathbb{R}_{\geq 0}$ for all $x\in \mathbb{R}_{\geq 0}^n$;
\item $\partial f_i/\partial x_i> 0$ and $\partial f_i/\partial x_j\geq 0$ for $i\neq j$;
\item $\alpha f_i(x) \leq f_i(\alpha x)$ for all $\alpha\in[0,1]$ and $x\in \mathbb{R}^n_{\geq 0}$;
\item $f_i(x)=x_i$ if $x_i=x_j$ for all $j\in\mathcal{N}_{i}^{in}$;
\item Inference graph $\mathcal{G}(f)$ has a globally reachable node;
\end{enumerate}
then, the MAS converges asymptotically to a consensus state for any initial state $x(0)\in\mathbb{R}_{\geq 0}^n$}.

\begin{pf*}{\textbf{Proof of Theorem \ref{th:nonlinearconsensus}}}
We start the proof by establishing the relations between properties $(i)-(v)$ and the following:
\begin{enumerate}[label=$(\alph*)$]
\item $f$ is positive;
\item $f$ is type-K order-preserving;
\item $f$ is sub-homogeneous;
\item $F_f=\{c\mathbf{1}: c\in \mathbb{R}_{\geq 0}\} $.
\end{enumerate}
We go through all equivalences one by one.
\begin{enumerate}

\item $[(i)\Leftrightarrow (a)]$ See Proof of Theorem \ref{th:nonlinearconvergence}.

\item $[(ii)\Leftrightarrow (b)]$ See Proposition \ref{th:ifflsoplike}.

\item $[(iii)\Leftrightarrow (c)]$ See Proof of Theorem \ref{th:nonlinearconvergence}.

\item $[(i-v)\Rightarrow (d)]$ The proof of this implication is given below.
\end{enumerate}

Condition $(iv)$ implies that the consensus space $c \mathbf{1}$ is a subset of the set of fixed points $F_f$ of map $f$, i.e., $$F_f\supseteq\{c\mathbf{1}: c\in \mathbb{R}_{\geq 0}\}.$$ 
%
%
By Theorem \ref{lem:stochasticJ} the Jacobian matrix $J_f(c\mathbf{1})$ evaluated at a consensus point is row-stochastic, i.e., $J_f(c\mathbf{1})\mathbf{1}=\mathbf{1}$. 
By the definition of inference graph (see Definition \ref{def:inferencegraph}), it holds that $\mathcal{G}(f)=\mathcal{G}(J_f(c\mathbf{1}))$. Thus $\mathcal{G}(J_f(c\mathbf{1}))$ has a globally reachable node by hypothesis and is aperiodic because condition $(ii)$ ensures a self-loop at each node. 

Now, we are ready to prove by contradiction that $[(i-v)\Rightarrow (d)]$. In particular, if there exists a fixed point $\bar{x}\neq c \mathbf{1}$, then by Lemma \ref{lem:invdir} the Jacobian matrix $J_f(c\mathbf{1})$ has a unitray eigenvalue with multiplicity strictly greater than one. On the other hand, by the widely known Theorem 5.1 in \cite{Bullo18}, if $\mathcal{G}(J_f(c\mathbf{1}))$ has a globally reachable node and is aperiodic then $J_f(c\mathbf{1})$ has a simple unitary eigenvalue with corresponding eigenvector equal to $\mathbf{1}$, unique up to a scaling factor $c$. 

This is a contradiction, therefore it does not exist a fixed point $\bar{x}$ such that $\bar{x}\neq c \mathbf{1}$ with $c> 0$. Thus, we conclude that the set of fixed points of map $f$ satisfies 
$$F_f = \{c\mathbf{1}, \ \ c\in \mathbb{R}_{\geq 0}\}.$$
Finally, if conditions $(a)$ to $(c)$ are satisfied, then Theorem \ref{th:nonlinearconvergence} the MAS converges to its set of fixed points $F_f$. If $(d)$ is satisfied, the $F_f$ contains only consensus points and thus the MAS in \eqref{eq:globdiscretesystem} converges to a consensus state for all $x\in\mathbb{R}^n_{\geq 0}$.
\hfill $\square$
\end{pf*}

\section{Examples} \label{sec:examples}

In this section we provide examples to corroborate our theoretical analysis of the convergences properties of discrete-time, nonlinear, positive, type-K order-preserving and sub-homogeneous MAS.

\begin{figure}[b!]
\begin{center}
 \begin{subfigure}{0.24\textwidth}
 \begin{center}
		\includegraphics[width=0.75\textwidth]{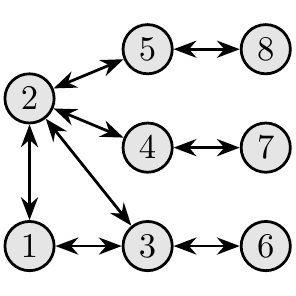}
		\caption{Graph $\mathcal{G}_1$.}\label{fig:graph_SIS}
		\end{center}
 \end{subfigure}%
 ~ 
 \begin{subfigure}{0.24\textwidth}
 \begin{center}
		\includegraphics[width=0.75\textwidth]{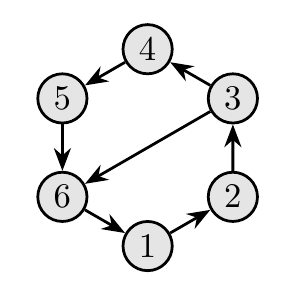}
		\caption{Graph $\mathcal{G}_2$.}\label{fig:graph_CON}
		\end{center}
 \end{subfigure}
 \caption{Graphs of Examples 1 and 2.}\label{fig:graphs}
\end{center}
\end{figure}

\begin{figure}[b!]
\begin{center}
 \begin{subfigure}[t]{0.25\textwidth}
 \begin{center}
		\includegraphics[width=\textwidth]{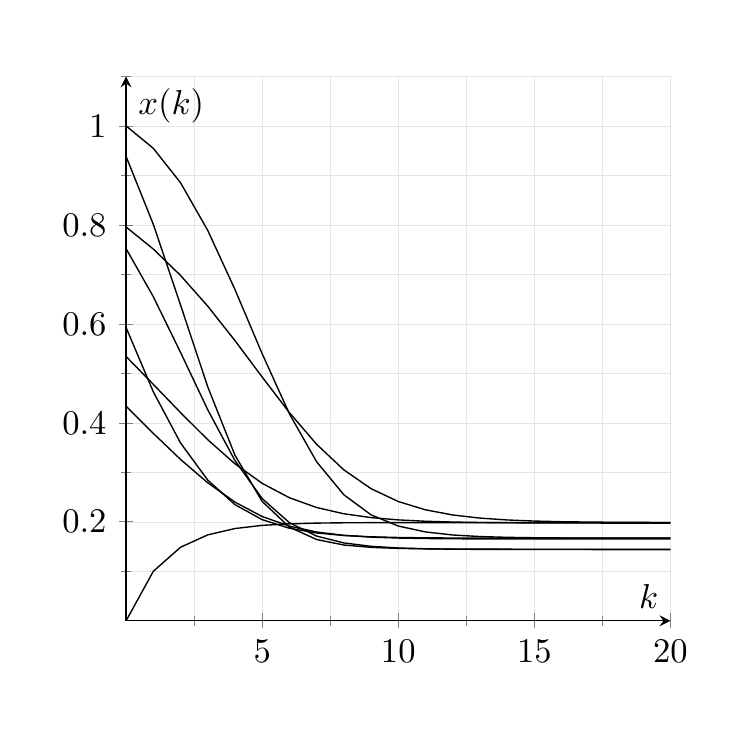}
		\caption{Parameters: $h=0.5$, $\beta_{ij} = 0.5$, $\delta_i=0.5$ for all $i$ and $j\in\mathcal{N}_i$.}\label{fig:ev_SIS}
		\end{center}
 \end{subfigure}%
 ~ 
 \begin{subfigure}[t]{0.25\textwidth}
 \begin{center}
		\includegraphics[width=\textwidth]{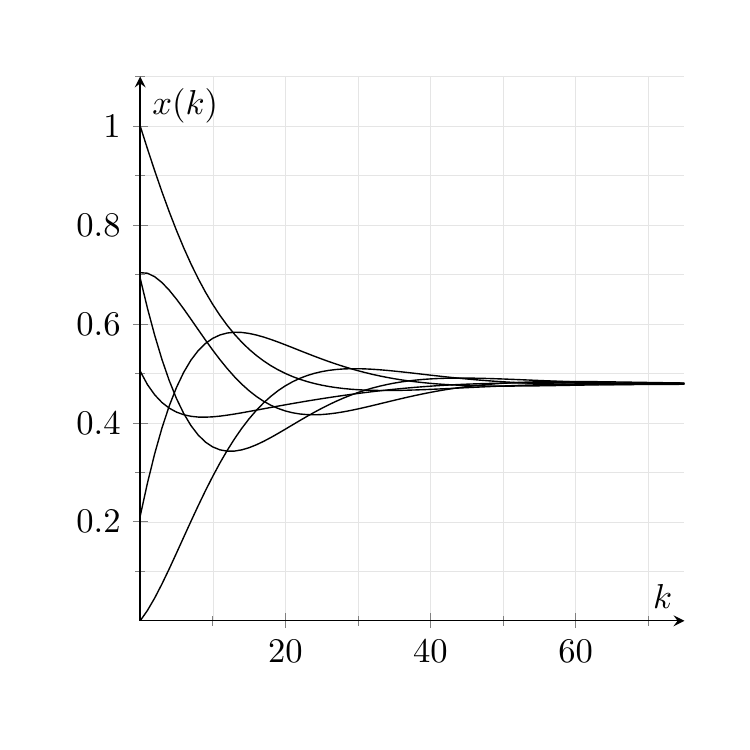}
		\caption{Parameters: $\varepsilon=0.1$.}\label{fig:ev_CON}
		\end{center}
 \end{subfigure}
 \caption{Evolutions of Examples 1 and 2.}\label{fig:evs}
\end{center}
\end{figure}

\begin{exmp} \normalfont
As a first example we consider a susceptible-infected-susceptible (SIS) epidemic model \cite{Allen94} described by the following 
\begin{align}
x_i(k+1)&=f_i(x(k))\label{eq:SISmodel} \\
& = \displaystyle x_i + h\left[\delta_i(1-x_i)-x_i\sum_{j\in\mathcal{N}_{i}}\beta_{ij}(1-x_j)\right]\:.\nonumber
\end{align}
Such model was originally derived to describe the propagation of an infectious diseases over a group of individuals. Each group is subdivided according to susceptible and infectious. Individuals can be cured and reinfected many times, there is not an immune group. Given $n$ groups, let $x_i(k)$, $y_i(k)$ be the portion of, respectively, susceptibles and infectious of group $i$ at time $k$, it is clear that $x_i(k),y_i(k)\geq 0$ and $x_i(k)+y_i(k)=1$ for any $k$. Thus, it is sufficient to consider the dynamics of one of them to completely describe the system. In model \eqref{eq:SISmodel} variables have the following meaning:
\begin{itemize}
\item $\beta_{ij}\geq 0$ are the infectious rates;
\item $\delta_i\geq 0$ is the healing rate;
\item $h\geq 0$ is the sampling rate.
\end{itemize}
We now evaluate conditions $(i)-(iv)$ of Theorem \ref{th:nonlinearconvergence} to establish the convergence of the associated MAS to a positive fixed point. Due to space limitations, we omit all steps and give directly conditions under which the theorem holds.
\begin{itemize}
\item First we notice that $x_i(k)$ belongs to $[0,1]$ for all $k$. It is guaranteed that for all $x_i(k)\in[0,1]$ also $x_i(k+1)+\in[0,1]$ if and only if \eqref{eq:pos1} and \eqref{eq:pos2} hold,
\begin{align}
&h\delta_i\leq 1\:,\quad h\sum_{j\neq i}\beta_{ij}\leq 1\:,\label{eq:pos1}\\
&h\beta_{ii}\leq\left[\sqrt{1-h\sum_{j\neq i}\beta_{ij}}+\sqrt{h\delta_i}\right]^2\:.\label{eq:pos2}
\end{align}
We conclude that for any $x\in [0,1]^n\subset \mathbb{R}_{\geq 0}^n$ then $f_i(x)\in[0,1]\subset \mathbb{R}_{\geq 0}$, thus proving that condition $(i)$ holds. 
\item Condition $(ii)$ holds if and only if the next inequality holds
\begin{equation}\label{eq:typek}
h\beta_{ii}< 1-h\delta_i-h\sum_{j\neq i}\beta_{ij}.
\end{equation}
\item Condition $(iii)$ holds if and only if the next inequality holds
\begin{equation}\label{eq:subhomo}
\delta_i\geq \sum_{j\neq i}\beta_{ij}.
\end{equation}
\item Condition $(iv)$ is satisfied since $\bar{x}=\mathbf{1}\in \mathbb{R}^n_{\geq 0}$ is a positive fixed point.
\end{itemize}
One can prove that \eqref{eq:pos1}, \eqref{eq:pos2}, \eqref{eq:typek}, \eqref{eq:subhomo} are equivalent to \eqref{eq:total}. Let $\beta = \sum_{j\neq i}\beta_{ij}$, 
\begin{align}\label{eq:total}
h\delta_i+h\beta< 1-h\beta_{ii}\:,\quad h\beta\leq 0.5\:.
\end{align}
If \eqref{eq:total} holds, then conditions of Theorem~\ref{th:nonlinearconvergence} are satisfied, and we conclude that the MAS converges to a fixed point for all $x\in[0,1]^n$. For a MAS described by graph $\mathcal{G}_1$ in Figure~\ref{fig:graph_SIS}, a numerical simulation is given in Figure~\ref{fig:ev_SIS}.

\end{exmp}

\begin{exmp} \label{ex:arctan} \normalfont Consider a MAS described by graph $\mathcal{G}_2$ in Figure~\ref{fig:graph_CON} and nonlinear local interaction rule
\begin{align}
x_i(k+1)&=f_i(x(k))\label{eq:CONmodel} \\
& = \displaystyle x_i(k)+\varepsilon_i \sum_{j\in\mathcal{N}_i} \text{atan}(x_j(k)-x_i(k))\:.\nonumber
\end{align}
We now evaluate conditions $(i)-(v)$ of Theorem \ref{th:nonlinearconsensus} to establish the convergence of the associated MAS to consensus state.
\begin{itemize}
\item Condition $(i)$ holds $\forall x\in\mathbb{R}^n_{\geq 0}$.
\item Condition $(ii)$ holds if and only if $\varepsilon_i \in \left(0,\left|\mathcal{N}_i\right|^{-1}\right]$ for $i=1,\ldots,n$.
\item Condition $(iii)$ holds $\forall x\in\mathbb{R}^n_{\geq 0}$.
\item Condition $(iv)$ is satisfied since $\bar{x}=c\mathbf{1}$ with $c\in\mathbb{R}_{\geq 0}$ is a solution for $x = f(x)$.
\item Condition $(v)$ is satisfied since graph $\mathcal{G}$ has a globally reachable node.
\end{itemize}
Thus, the conditions of Theorem~\ref{th:nonlinearconsensus} are satisfied, and we conclude that the MAS in \eqref{eq:CONmodel} converges to a consensus state. A numerical simulation is given in Figure~\ref{fig:ev_CON}.
\end{exmp}

\section{Conclusions and future works} \label{conclusion}

In this paper we presented three main results related to a class of nonlinear discrete-time multi-agent systems represented by a state transition map which is positive, sub-homogeneous and type-K order preserving.

	The first result establishes that a general discrete-time dynamical system converges to one of its equilibrium points asymptotically if its corresponding state transition map is positive, sub-homogeneous and type-K order preserving.
	The second result provides sufficient conditions for a set of nonlinear, discrete-time heterogeneous local interaction rules which define the MAS to establish stability of the MAS, independently of its graph topology (which is considered unknown) by exploiting our first main result.
	Finally, the third result provides sufficient conditions for a set of nonlinear, discrete-time heterogeneous local interaction rules which define the MAS to establish asymptotic convergence to a consensus state if the inference graph of the MAS has a globally reachable node.

This paper generalizes results for discrete-time linear MAS whose state transition matrix is stochastic to the nonlinear case thanks to nonlinear Perron-Frobenius theory. Examples are provided to show the effectiveness of the stability analysis of a MAS based on our method.

Future work will consider MAS represented by a time-varying set of heterogeneous local interaction rules.




%


\bibliographystyle{plain} 
\bibliography{autosam} 

\begin{thebibliography}{10}

\bibitem{GaubertAkian2006}
M.~Akian, S.~Gaubert, B.~Lemmens, and R.~Nussbaum.
\newblock Iteration of order preserving subhomogeneous maps on a cone.
\newblock {\em Mathematical Proceedings of the Cambridge Philosophical
  Society}, 140(1):157--176, 2006.

\bibitem{GaubertAkian2003}
Marianne Akian and St{\'e}phane Gaubert.
\newblock Spectral theorem for convex monotone homogeneous maps, and ergodic
  control.
\newblock {\em Nonlinear Analysis: Theory, Methods {\&} Applications},
  52(2):637 -- 679, 2003.

\bibitem{Allen94}
Linda J.~S. Allen.
\newblock Some discrete-time si, sir, and sis epidemic models.
\newblock {\em Mathematical biosciences}, 124 1:83--105, 1994.

\bibitem{Bullo18}
F.~Bullo.
\newblock {\em Lectures on Network Systems}.
\newblock Version 0.96, 2018.
\newblock With contributions by J. Cortes, F. Dorfler, and S. Martinez.

\bibitem{Deplano18}
D.~{Deplano}, M.~{Franceschelli}, and A.~{Giua}.
\newblock Lyapunov-free analysis for consensus of nonlinear discrete- time
  multi-agent systems.
\newblock In {\em 2018 IEEE Conference on Decision and Control (CDC)}, pages
  2525--2530, Dec 2018.

\bibitem{Fran2017}
M.~Franceschelli, A.~Giua, and A.~Pisano.
\newblock Finite-time consensus on the median value with robustness properties.
\newblock {\em IEEE Transactions on Automatic Control}, 62(4):1652--1667, April
  2017.

\bibitem{Fran2015}
M.~Franceschelli, A.~Pisano, A.~Giua, and E.~Usai.
\newblock Finite-time consensus with disturbance rejection by discontinuous
  local interactions in directed graphs.
\newblock {\em IEEE Transactions on Automatic Control}, 60(4):1133--1138, April
  2015.

\bibitem{Gunawardena2003}
Jeremy Gunawardena.
\newblock From max-plus algebra to nonexpansive mappings: a nonlinear theory
  for discrete event systems.
\newblock {\em Theoretical Computer Science}, 293(1):141 -- 167, 2003.

\bibitem{Hirsch2006}
M.W. Hirsch and Hal Smith.
\newblock Chapter 4 monotone dynamical systems.
\newblock volume~2 of {\em Handbook of Differential Equations: Ordinary
  Differential Equations}, pages 239 -- 357. North-Holland, 2006.

\bibitem{jadbabaie2003coordination}
A.~Jadbabaie, J.~Lin, and A.S. Morse.
\newblock Coordination of groups of mobile autonomous agents using nearest
  neighbor rules.
\newblock {\em IEEE Transactions on Automatic Control}, 48(6):988--1001, 2003.

\bibitem{Jiang1996}
J.~F. Jiang.
\newblock Sublinear discrete-time order-preserving dynamical systems.
\newblock {\em Mathematical Proceedings of the Cambridge Philosophical
  Society}, 119(3):561--574, 1996.

\bibitem{Kamke32}
Erich Kamke.
\newblock Zur theorie der systeme gewöhnlicher differentialgleichungen. ii.
\newblock {\em Acta Mathematica 58}, pages 57--85, 1932.

\bibitem{Lemmens2006nlp}
Bas Lemmens.
\newblock Nonlinear perron-frobenius theory and dynamics of cone maps.
\newblock {\em Positive systems}, pages 399--406, 2006.

\bibitem{LemmensNussbaum2012}
Bas Lemmens and Roger Nussbaum.
\newblock {\em Nonlinear Perron-Frobenius Theory}.
\newblock Cambridge Tracts in Mathematics. Cambridge University Press, 2012.

\bibitem{Lemmens2005}
Bas Lemmens and Michael Scheutzow.
\newblock On the dynamics of sup-norm non-expansive maps.
\newblock {\em Ergodic Theory and Dynamical Systems}, 25(3):861--871, 2005.

\bibitem{Liu13}
Yang-Yu Liu, Jean-Jacques Slotine, and Albert-L{\'a}szl{\'o} Barab{\'a}si.
\newblock Observability of complex systems.
\newblock {\em Proceedings of the National Academy of Sciences},
  110(7):2460--2465, 2013.

\bibitem{Moreau05}
L.~{Moreau}.
\newblock Stability of multiagent systems with time-dependent communication
  links.
\newblock {\em IEEE Transactions on Automatic Control}, 50(2):169--182, Feb
  2005.

\bibitem{Nussbaum1990}
Roger~D. Nussbaum.
\newblock Omega limit sets of nonexpansive maps: finiteness and cardinality
  estimates.
\newblock {\em Differential Integral Equations}, 3(3):523--540, 1990.

\bibitem{Olfati2007}
R.~Olfati-Saber, J.~A. Fax, and R.~M. Murray.
\newblock Consensus and cooperation in networked multi-agent systems.
\newblock {\em Proceedings of the IEEE}, 95(1):215--233, Jan 2007.

\bibitem{Olshevsky2008}
A.~Olshevsky and J.~N. Tsitsiklis.
\newblock On the nonexistence of quadratic lyapunov functions for consensus
  algorithms.
\newblock {\em IEEE Transactions on Automatic Control}, 53(11):2642--2645, Dec
  2008.

\bibitem{Polacik1992}
P.~Pol{\'a}{\v{c}}ik and I.~Tere{\v{s}}{\v{c}}{\'a}k.
\newblock Convergence to cycles as a typical asymptotic behavior in smooth
  strongly monotone discrete-time dynamical systems.
\newblock {\em Archive for Rational Mechanics and Analysis}, 116(4):339--360,
  Dec 1992.

\bibitem{Valcher18}
Anders Rantzer and Maria~Elena Valcher.
\newblock A tutorial on positive systems and large scale control.
\newblock pages 3686--3697, 12 2018.

\bibitem{Rosenberg2001}
Dinah Rosenberg and Sylvain Sorin.
\newblock An operator approach to zero-sum repeated games.
\newblock {\em Israel Journal of Mathematics}, 121(1):221--246, Dec 2001.

\bibitem{Smith88}
Hal~L. Smith.
\newblock Systems of ordinary differential equations which generate an order
  preserving flow. a survey of results.
\newblock {\em SIAM Review}, 30(1):87--113, 1988.

\bibitem{Sun2009}
Y.~G. Sun and L.~Wang.
\newblock Consensus of multi-agent systems in directed networks with nonuniform
  time-varying delays.
\newblock {\em IEEE Transactions on Automatic Control}, 54(7):1607--1613, July
  2009.

\bibitem{Thompson64}
A.~C. Thompson.
\newblock On certain contraction mappings in a partially ordered vector space.
\newblock {\em Proceedings of the American Mathematical Society},
  14(3):438--443, 1963.

\bibitem{ZhiyunLin2007}
L.~Zhiyun, B.~Francis, and Maggiore M.
\newblock State agreement for continuous-time coupled nonlinear systems.
\newblock {\em SIAM Journal on Control and Optimization}, 46(1):288--307, 2007.

\end{thebibliography}

\appendix
\end{document}